\documentclass[12pt,preprint]{aastex}

\newcommand{\degg}{\hbox{$^\circ$}}
\newcommand{\PG}{PG\,1402+261}
\newcommand{\xmm}{{\it XMM-Newton}}
\newcommand{\asca}{{\it ASCA}}
\newcommand{\sax}{{\it Beppo-SAX}}
\newcommand{\arcm}{\hbox{$^\prime$}}
\newcommand{\arcs}{\hbox{$^{\prime\prime}$}}

\newcommand{\ls}
{\mathrel{\hbox{\rlap{\hbox{\lower4pt\hbox{$\sim$}}}\hbox{$<$}}}}
\newcommand{\gs}
{\mathrel{\hbox{\rlap{\hbox{\lower4pt\hbox{$\sim$}}}\hbox{$>$}}}}

\received{}
\begin{document}
\title{An extreme, blueshifted iron line profile in the Narrow Line Seyfert 1 
PG\,1402+261; an edge-on accretion disk or highly ionized absorption?}
\shorttitle{The XMM-Newton observation of the NLS1 PG\,1402+261}
\shortauthors{Reeves \et}
\author{ J.N. Reeves\altaffilmark{1,2}, D. Porquet\altaffilmark{3}, T.J. Turner\altaffilmark{1,4}}
\email{jnr@milkyway.gsfc.nasa.gov, dporquet@mpe.mpg.de, 
turner@milkyway.gsfc.nasa.gov}

\altaffiltext{1}{Laboratory for High Energy Astrophysics, Code 662, 
NASA Goddard Space Flight Center, Greenbelt Road, Greenbelt, MD 20771, USA}
\altaffiltext{2}{Universities Space Research Association, 7501 Forbes 
Boulevard, Suite 206, Seabrook, MD 20706, USA}
\altaffiltext{3}{Max-Plank-Institut f{\" u}r extraterrestrische Physik, Postfach 1312, 85741 Garching, Germany}
\altaffiltext{4}{Joint Center for Astrophysics, University of Maryland 
Baltimore County, 1000 Hilltop Circle, Baltimore, MD 21250, USA}

\begin{abstract}

We report on a short \xmm\ observation of the radio-quiet Narrow Line 
Seyfert 1 \PG. 
The EPIC X-ray spectrum of \PG\ shows a strong 
excess of counts between $6-9$~keV in the rest frame. This feature can be 
modeled by an unusually strong (equivalent width 2~keV) and 
very broad (FWHM velocity of 110000~km~s$^{-1}$) 
iron K-shell emission line. The line centroid energy at 7.3 keV 
appears blue-shifted with respect to the iron 
K$\alpha$ emission band between $6.4-6.97$~keV, while the blue-wing of the 
line extends to 9~keV in the quasar rest frame.  
The line profile can be fitted 
by reflection from the inner accretion disk, but an inclination angle of 
$>60\degg$ is 
required to model the extreme blue-wing of the line. Furthermore the 
extreme strength of the line requires a geometry whereby the hard X-ray 
emission from \PG\ above 2 keV is dominated by the pure-reflection component 
from the disk, while little or none of the direct hard power-law is observed. 
Alternatively the spectrum above 2 keV may instead be explained by an ionized 
absorber, if the column density is sufficiently high 
($N_{\rm H}>3\times10^{23}$~cm$^{-2}$) and if the matter 
is ionized enough to produce a deep ($\tau\sim1$) iron K-shell 
absorption edge at 9 keV. This absorber could originate in 
a large column density, high velocity outflow, perhaps similar to those 
which appear to be observed in several other high accretion rate AGN. 
Further observations, especially at higher spectral resolution, 
are required to distinguish between the accretion disk reflection 
or outflow scenarios.

\end{abstract}


\section{Introduction}

\PG\ is a type-I radio-quiet quasar at z=0.164 \citep{Schmidt83}.  
It emits narrow optical permitted lines with FHWM 
(H$\beta$)=1910\,km\,s$^{-1}$ 
and can also be classified as a Narrow Line Seyfert 1 (NLS1), 
which are defined to have FWHM (H$\beta$)$\,<2000$\,km\,s$^{-1}$ 
\citep{Oster87}. 
The mass of the supermassive black hole in this quasar has been 
estimated to be 2$\times$10$^{7}$\,M$_{\odot}$ 
and the bolometric luminosity is about 1.3$\times$10$^{45}$\,erg\,s$^{-1}$
\citep{Woo02}.
This leads to a high accretion rate relative to Eddington 
of about 50$\%$ \citep{PROB04}. \PG\ was previously observed in X-rays with
\sax\ by \cite{M00}.
They found that the data were satisfactorily represented by
a broken power law continuum of $\Gamma=2.59\pm0.10$ 
and $\Gamma=1.52\pm0.30$ for the soft and hard X-ray bands, respectively, 
although there was some evidence for an excess of counts above 7~keV in the 
MECS spectrum. 
A preliminary analysis of a recent (2002) XMM-Newton observation of \PG, 
reported by \cite{PROB04} among a sample of PG quasars, 
shows that the spectrum cannot be represented by a simple 
broken power-law continuum alone, 
but exhibits a large deviation between $6-9$~keV in the quasar rest frame, 
in the form of an apparent broad and blue-shifted iron K emission line.
This observation adds to the number of AGN (and NLS1s in particular) 
that appear to exhibit unusually strong emission 
or absorption features above 7 keV in the iron K-shell band; e.g. 1H~0707-495 
\citep{Boller02}, IRAS~13224-3809 \citep{Boller03}, PDS~456 
\citep{R03} and PG~1211+143 \citep{Pounds03}. 
These iron K band features have either been interpreted as the signature of  
enhanced reflection from the inner accretion disk, e.g. 
1H~0707-495 \citep{Fabian04} but 
in some cases may also be interpreted as evidence for an additional 
high column density absorber, either from matter 
partially covering the X-ray source \citep{Gallo04, Tanaka04} 
or even from a high velocity ($\sim0.1c$) 
outflow \citep{Pounds03, R03}.   


Here we present a spectral analysis of the \xmm\ observation of 
\PG, modeling the strong iron K-band emission (or absorption) feature.  
We investigate the possible physical origins for this feature; either in the 
form of an extreme broadened iron emission line produced via 
X-ray reflection from accretion disk or alternatively 
a very high column density ionized absorber. 
Note that all fit parameters are given in the quasar rest frame,
with values of H$_{\rm 0}$=75\,km\,s$^{-1}$\,Mpc$^{-1}$,
and q$_{\rm 0}$=0.5 assumed throughout. Errors are quoted at 90\% 
confidence (e.g. $\Delta\chi^{2}=2.7$, for 1 parameter of interest). 

\section{The XMM-Newton observation}
 
\PG\ was observed by \xmm\ on 2002 February 27 (OBSID: 0109081001),  
with an exposure time of 
only 9.1\,ks. Data were taken with the EPIC pn and 
MOS CCDs \citep{Struder01,Turner01},   
in Large Window Mode with the thin filter. The data was
reduced using version 5.4.1 of the XMM-SAS software.
Data were selected using event patterns 0-4 and 0-12 for pn and MOS 
respectively, whilst only good X-ray events 
(with `FLAG=0') were included. The source spectra were extracted from a 
circular source region of $\sim40$\arcs\ radius, whilst background spectra were
extracted from a box of 1.85\arcm\ radius around the source, but excluding 
the circular source region. 
Response matrices and ancillary response
files were generated using the SAS tasks \textsc{rmfgen} and 
\textsc{arfgen} respectively. EPIC-pn lightcurves were also extracted 
over two energy bands, however no dramatic variability was observed, 
therefore the spectral analysis is performed on the whole \xmm\ exposure. 

\section{X-ray spectral analysis}

In all subsequent fits, the column density  is fixed to the 
Galactic value, i.e, 1.47 $\times$ 10$^{20}$\,cm$^{-2}$. Both the EPIC-pn 
and the co-added MOS spectra are shown in Figure~1, with the data/model ratio 
residuals plotted against a power-law continuum of photon index 
$\Gamma=2.2$ fitted in the $2-5$~keV band. A 
strong soft excess is present below 2 keV and the continuum shape can be 
parameterized by a simple broken power-law, with a soft index of 
$\Gamma=2.84\pm0.04$ and a hard index of $\Gamma=2.17\pm0.12$ 
above a break energy of $E=1.8\pm0.2$~keV. 
A hard excess is also clearly observed above 5 keV in both the 
pn and MOS detectors. 
For this broken-powerlaw continuum, 
the unabsorbed 0.3--10\,keV (2--10\,keV) band flux is  
7.3$\times$10$^{-12}$\,erg\,cm$^{-2}$\,s$^{-1}$
(1.8$\times$10$^{-12}$\,erg\,cm$^{-2}$\,s$^{-1}$),  
with a corresponding luminosity of 4.4$\times$10$^{44}$\,erg\,s$^{-1}$ 
(1.0$\times$10$^{44}$\,erg\,s$^{-1}$). About 77$\%$ of the flux is emitted 
below 2\,keV, consistent with the strong soft X-ray excess. Note that at this 
flux level, photon pile-up of the source X-ray spectrum is negligible. \\

As the pn spectrum contains significantly more counts than the 
MOS above 6 keV in the iron K band, we now concentrate on the pn 
spectral analysis. Figure 2 shows the 
data/model ratio to the broken-power-law continuum, plotted from 3-11 keV 
in the quasar rest frame. Notice that the majority of the flux of the iron 
K band feature is observed above 7 keV, i.e. above the rest energy for H-like 
iron, whilst a sharp drop in the spectrum is observed at 9 keV in the rest 
frame. A single broad Gaussian emission line models the hard excess rather 
well, with a centroid energy of 
${\rm E}=7.3^{+0.4}_{-0.5}$~keV, a line width of 
$\sigma=1.2^{+0.6}_{-0.4}$~keV (FWHM velocity of 
$1.1\times10^{5}$~km~s$^{-1}$) and a very high equivalent width of 
$\sim2$~keV. 
The fit statistic is acceptable, with $\chi^2/{\rm dof}=390.5/377$ (where 
dof is the number od degrees of freedom), 
with the feature detected at $>99.99\%$ significance according to 
an F-test, whilst the line centroid is formally consistent with emission   
from hydrogen like iron (at 6.97 keV). 

It might be reasonable to expect a strong reflected continuum to accompany 
the iron line in \PG. Therefore we performed a spectral fit 
using an X-ray continuum model based on ionized reflection from the inner 
accretion disk, using the \textsc{pexriv} model within \textsc{xspec} 
\citep{Mag95}. We assume a disk 
inclination angle of 60 degrees 
(see section 3.2 for more detailed reflection fits), 
a disk temperature of $10^6$~K 
and use solar abundances, which are kept fixed in the spectral fit. 
A broad Gaussian emission line is also included in the fit, 
as the \textsc{pexriv} model contains only the 
reflected continuum and not the iron emission line. The reflected continuum 
is smoothed with a Gaussian function, identical in width 
to the broad iron line. The strength of the reflector $R$  
is conventionally defined as the ratio of the solid angle subtended 
by the reflecting material to $2\pi$~steradian, whilst the ionization 
parameter is defined as $\xi=4\pi F_{\rm ion}/n$, where n is the 
density of the reflector and $F_{\rm ion}$ is the ionizing flux integrated 
over the 
5~eV to 20~keV band. The fits results are shown in Table 1, fit 1. The 
reflection model provides an acceptable fit with $R=1.6\pm0.7$ and a 
relatively high ionization parameter of log~$\xi=3.2\pm0.3$~erg~cm~s$^{-1}$. 
In this model, the continuum emission in \PG\ is parameterized with a 
single, steep power-law of $\Gamma=2.47\pm0.08$, 
whilst the highly ionized matter is sufficiently reflective 
at lower energies to model the soft excess below 2 keV. 
Note that the reflection  
fits are largely insensitive to the iron abundance, as the iron K shell edge  
present in the continuum reflection spectrum is highly smeared. 
Even upon the addition of the reflector, the properties of the iron line 
are largely unchanged, with the equivalent width now being slightly 
lower at $\sim1.5$~keV. 

In Figure~2, it appears that there may be two peaks within the \xmm\ 
line profile of \PG, one near 7.0--7.5\,keV and the other near 8.5\,keV. 
For instance one line may originate from Fe~\textsc{xxvi}~K$\alpha$ 
(at 7.0~keV), the other peak from Fe~\textsc{xxvi}~K$\beta$ or 
Ni~\textsc{xxviii}~K$\alpha$ (near 8.0 keV). To parameterize this, 
we fitted the iron line profile of \PG\ with two Gaussian lines, 
one with a rest energy of $7.1\pm0.3$~keV (i.e. consistent with
Fe~\textsc{xxvi} K$\alpha$) and another line at $\sim8.6\pm0.4$~keV. 
For this fit, we revert to the simple broken power-law 
continuum (without reflection), as described above. 
The equivalent width of the line at 
7.1 keV line is much stronger than the higher energy line (1.3\,keV vs 
0.4~keV). The fit is only slightly improved for the 
addition of three extra parameters compared 
to a single Gaussian line fit ($\chi^{2}/{\rm dof}=387.4/374$ vs 390.5/377), 
therefore we consider the 2 Gaussian line fit to be an over-parameterization 
of the data. Nonetheless 
we do not rule out an additional contribution towards the line from 
either Fe K$\beta$ or Ni K$\alpha$. A further, much longer observation 
will be required with \xmm\ in order to investigate 
structure within the line profile. 

\subsection{Comparison with a previous \asca\ observation.}

\PG\ was also previously observed by \asca\ on January 12, 2000.
The \asca\ data are composed of the SIS0 (exposure time $\sim$ 43.2\,ks), 
SIS1 ($\sim$ 37.7\,ks), GIS2 and GIS3 (each $\sim$ 57.3\,ks) data.
The background subtracted source spectra were subsequently analyzed, 
using a minimum of 20 counts per spectral bin. 
Figure 3 shows the data/model ratio of the GIS and SIS data from $1-10$\,keV 
to a simple power-law model with Galactic absorption, plotted in the 
quasar rest frame. 
There is a clear ``line-like'' excess in the data between 6--8.5~keV 
in the rest frame, similar to the residuals observed in the \xmm\ spectrum of 
\PG. Subsequently, a fit above 1~keV using all four detectors, 
with a single power-law ($\Gamma=2.3\pm0.2$)  
plus a broad emission Gaussian line,  
gives a reasonable representation of the data 
($\chi^{2}$/d.o.f=404.2/361, null hypothesis probability of $5\times10^{-2}$) 
and the line is required at a confidence level of $>$99.95$\%$. 
 The energy of the line is  7.2$^{+0.4}_{-0.6}$\,keV
 and its width is high with $\sigma$=0.8$^{+0.7}_{-0.4}$\,keV. The 
energy of the line is consistent with emission from He-like or H-like iron 
(at 6.7 and 6.97~keV respectively). 
Note that there are no other apparent residuals present in the 
spectrum below the iron K line energy. 
The EW of $\sim1$~keV is lower than found during
 the XMM-Newton observation, but is compatible within the errors. 
 During the ASCA observation the 2--10\,keV unabsorbed flux was
 2.5$\times$10$^{-12}$\,erg\,cm$^{-2}$\,s$^{-1}$ (SIS0), 
i.e. about 40$\%$ higher than the XMM-Newton observation. 

The addition of a reflection component  
to the model (the \textsc{pexriv} model as described above) 
does not strongly effect the line properties 
(see Table 1, fit 2). As the constraints on the reflected continuum from the 
\asca\ data are worse than for \xmm, the ionization parameter was 
fixed at log~$\xi=3.0$~erg~cm~s$^{-1}$, whilst R was fixed at 2 (i.e. the 
maximum permitted value, corresponding to $4\pi$~steradian solid angle). 
The effect on the line is to slightly reduce its equivalent
width, whilst the line width is then poorly constrained ($\sigma<1.4$~keV) 
when the reflected continuum is included in the model. 

\subsection{A broad disk emission line?}

Given the large width of the emission line in the \xmm\ (and \asca) fits, we 
attempted to fit the iron line profile with emission from the inner accretion 
disk around a Kerr black hole, using the xspec \textsc{laor} 
model \citep{Laor91}. An ionized reflection component was also 
included in the fits, in order to model the reflected continuum emission 
from the surface of the accretion disk. The reflected continuum 
is relativistically smeared using identical parameters as 
in the disk emission line model. 
We assume an inner radius of $1.2R_{g}$ ($R_{g}=GM/c^{2}$, i.e. one  
gravitational radius) and an outer radius of $400R_{g}$. Initially we fix 
the disk inclination angle at $30\degg$, reasonable for a type I AGN. 
We obtain a very high line energy of 8.5$^{+0.4}_{-0.3}$\,keV, required 
to fit the blue-wing of the line between $8.5-9$~keV, whilst the 
line equivalent width is extremely large at $\sim4$~keV, 
relative to the power-law continuum at 8.7~keV (Table 1, fit 3). 
The line energy 
is considerably larger than the rest energy of H-like iron
(6.97\,keV), 
therefore this could indicate a significant outflow of the medium producing
 the line, i.e. 0.3--0.4\,c (90\,000--120\,000\,km\,s$^{-1}$), 
for H-like or neutral iron respectively. As an alternative model, 
we free the disk inclination angle, but instead fix the rest energy of the 
iron line emission at 6.97 keV, i.e. corresponding to H-like iron. 
This then requires a 
large disk inclination of $68\pm5$ degrees in order to fit the 
blue-wing of the 
line, although the fit is equally acceptable (Table~1, fit 4), 
with an equivalent width of $\sim1.7$~keV.    
Thus either a substantial outflow of the line emitting matter is required, 
or the disk has to be highly inclined to the line of sight. 

 
Given the evidence for a very strong broad line in 
\PG, we tested whether the spectrum and line profile 
could be fitted with an 
ionized disk reflection model. 
We use the model {\sc xion} from \cite{Nay00} in the simplest lamppost 
configuration to model the emission above 2~keV, 
together with a soft featureless power-law component to 
model the steep soft X-ray excess in \PG. 
We fixed the height of the source above the disk to $10R_{g}$, the 
inner and outer disk radii to $4R_{g}$ (the minimum the model allows) and 
$100R_{g}$ respectively. The accretion rate 
was fixed to 50\% of Eddington, whilst the ratio of X-ray to disc (UV) 
luminosity was set to 10\% \citep{PROB04}. 
We find the reflection model fits the spectrum well 
with $\Gamma_{\rm soft}$=3.11$^{+0.09}_{-0.08}$ and  
 $\Gamma_{\rm hard}$=1.70$^{+0.17}_{-0.44}$ for the continuum photon 
indices and 
a fit statistic of $\chi^{2}/dof=397/379$. The unfolded spectral fit is 
shown in Figure 4. Although the model produces 
mainly He and H-like Fe emission (at 6.7~keV and 6.97~keV respectively), 
a high disk inclination of 
greater than $>$60\degg\ is required to model the extreme blue-shift of 
the line. In addition, a high iron abundance of $5\times$ solar (the 
maximum permitted value) is required, whilst in order to model the 
extreme strength of the line, the hard emission 
component is required to be {\it reflection dominated},  
i.e. the illuminating hard power-law component cannot be directly observed. 

As a consistency check, we compare the reflection results 
obtained with the \textsc{xion} model with the  
predictions of \citet{Btyne02} who compute reflection spectra from a 
constant density slab, illuminated by a power-law continuum. 
In their model, for 
pure reflected emission (i.e. with no direct power-law observed), 
a line equivalent width of $\sim$1.5\,keV in \PG\ requires 
an iron over-abundance of $>3\times$ solar, assuming a reflection ionization 
parameter of log~$\xi\sim3-4$ and an illuminating power-law continuum of 
$\Gamma=2.3$. This appears to be consistent with the reflection fits presented 
above. In the scenario where both the primary power-law and reflected 
emission are both observed (corresponding to the reflector covering 
$2\pi$~steradian solid angle), then the predicted line is not strong enough
to account for the iron line feature in \PG. Even for a $10\times$ 
solar abundance of iron, the maximum predicted 
equivalent width is $\sim700$\,eV. 
Thus it is likely that the X-ray emission from 
\PG\ would have to be reflection dominated to explain the 
apparent strong iron line in the \xmm\ data. 

\subsection{Highly Ionized Absorption in \PG?}

Alternatively, we have investigated whether the 
strong iron K feature can be modeled with a highly ionized absorber. 
One can fit the spectrum reasonably well with a continuum modeled by two 
power-laws ($\Gamma_{\rm soft}=3.0\pm0.2$ and $\Gamma_{\rm hard}=0.9\pm0.5$), 
although the hard power-law has to be very flat to model the hard excess 
above 5~keV. The addition of an absorption edge at $9.0\pm0.2$~keV, with a 
depth of $\tau=1.1\pm0.5$, models the sharp 
drop at 9 keV in the rest frame 
(Figure 2), with a fit statistic of $\chi^{2}/dof=395/378$. The energy of 
the K-shell edge then either corresponds to  
H-like iron (at 9.27 keV) or He-like iron 
(at 8.75 keV). 

We then proceed to fit the data with the \textsc{xspec} warm absorber model 
\textsc{absori} \citep{Done92}. 
The model is applied so that the soft X-ray power-law 
(which may for instance originate directly from the disk) 
is unattenuated, but so 
the hard X-ray power-law is modified by the absorber 
(i.e. a partial coverer). In order to model the depth of the edge
a large column density of $N_{H}\sim 2\times10^{24}$~cm$^{-2}$ is required. 
Formally, the 90\% confidence lower-limit obtained on the ionization parameter 
from this model is log~$\xi>3.5$~erg~cm~s$^{-1}$, with 
an outflow velocity of $<20000$~km~s$^{-1}$, whilst the column density 
must exceed $N_{\rm H}>3\times10^{23}$~cm$^{-2}$ 
(assuming $5\times$ solar abundance). 
The best fit model is shown in Figure 5, whilst the fit obtained is equally as 
good as the disk reflection model 
($\chi^{2}/{\rm dof}=396/378$). Inclusion of the warm absorber 
results in there now only being 
a requirement for a {\it narrow} iron K$\alpha$ 
emission line, with an equivalent width of $\sim600$~eV. The line 
centroid energy is $7.4\pm0.1$~keV, 
which indicates that there may still be some 
blue-shift of this component; e.g., $\sim18000$~km~s$^{-1}$ compared to the 
rest-frame energy of H-like Fe at 6.97 keV. 

The partial covering model can also be applied to the \asca\ spectrum of 
\PG. The fit obtained is equally good as the broad line fit reported in 
section 3.1 (fit statistic $\chi^{2}/{\rm dof}=404.2/360$). As above, the 
partial absorber covering is applied to the hard power-law, whilst a 
column density of $N_{H}\sim1.2\times10^{24}$~cm$^{-2}$ and an ionization 
parameter of log~$\xi\sim3.3$~erg~cm~s$^{-1}$ are required, similar to the 
values obtained above.

\section{Discussion and Conclusions}

The X-ray spectrum of \PG\ shows a strong excess in the 
iron K-shell band between $6-9$~keV. 
If this is due to a relativistic disk line, then its parameters 
are rather extreme. The line equivalent width measured by \xmm\ is $>2$~keV, 
even stronger than the broad iron line observed in MCG~-6-30-15 
\citep{Tanaka95, Wilms01, VF04}. 
Such a strong iron K$\alpha$ emission line can only be produced if the 
hard X-ray emission in \PG\ is dominated by the disk reflection component, 
with little of the direct hard power-law being 
observed. Furthermore the rest-energy of the line, peaking above 7 keV 
and extending up to 9 keV, implies that the accretion 
disk is likely to be observed at large inclination angles, i.e. $>60\degg$. 
Such a high inclination is unusual 
for a type I AGN in the context of AGN unification schemes 
\citep{Anton93}. Interestingly a high (50 degrees) inclination 
was recently inferred from the reflection fits to the 
NLS1 galaxy, 1H\,0707-495 \citep{Fabian04}. Taken together, the high 
inclination angle required in the reflection fits 
for both \PG\ and 1H\,0707-495 may argue 
against the pole-on model for some NLS1s \citep{Bian04}.

It is possible that large density perturbations  
in the disk could explain the strong 
reflection dominated emission above 2~keV.  
Variations of up to $\times100$ in density 
have predicted from simulations of highly turbulent disks 
\citep{NTurner02}, whilst reflection spectra have been 
calculated for inhomogeneous disks \citep{Btyne04}. Thus a 
natural geometry could occur 
whereby the direct hard X-ray emission (e.g. from compact 
coronal flares) is obscured by dense, turbulent clouds within the  
disk surface layers, the same clouds 
then reflect the X-rays into the line of sight \citep{GR88}. 
A similar geometry was 
applied by \cite{Fabian02} to the Narrow Line Seyfert 1, 
1H~0707-495, which shows an unusually strong iron K edge at 7.1 keV.  
\cite{Fabian02} propose that the hard X-rays could be emitted within 
low density cavities between 
dense rings of matter on the disk surface; the ring surfaces then 
reflect the X-rays into the observers line of sight, whilst the 
direct emission is obscured, producing a strong iron K line. 
In \PG, the direct X-rays are more likely to be obscured
if the disk was viewed at large (side-on) inclination angles, 
as is inferred from our accretion disk fits. The soft X-ray 
(and EUV) excess emission may not be obscured however,  
if this is the Comptonized thermal emission from the 
extended inner disk region \citep{Malkan82, Czerny87}, although the 
enhanced (reflected) emission from the surface of the ionized disk 
may also contribute towards the soft X-ray excess.  

In the inhomogeneous reflection models of Ballantyne, Turner \& Blaes (2004), 
the authors find that it may be difficult to produce a very strong 
iron line. However this is dependent on the geometry, if the reflected 
X-rays undergo multiple scattering (i.e. the reflected X-rays are themselves 
reflected), then the resultant line equivalent width can be very high 
\citep{Ross02}, as large as 6\,keV. 

One other possibility is that the disk reflection is enhanced by the 
effects of gravitational light bending of the hard X-ray emission 
{\it towards} the disk, i.e. returning radiation \citep{Cunn75}. 
This effect will be strongest when the X-rays are 
concentrated within $2-3 R_{g}$ around 
the spin axis of a maximally rotating black hole 
(in the so-called ''lamppost'' configuration). Indeed it has been 
suggested that the strong, redshifted iron line in MCG~-6-30-15 may be 
boosted by this process \citep{Mart02, FV02}. However in order 
to explain the observed blueshift of the iron line profile, 
especially as a significant amount of the line flux would be gravitationally 
redshifted below 6\,keV in the above scenario, then the disk 
inclination angle in \PG\ must be nearly side-on.

Alternatively, the spectrum of \PG\ can also 
be modeled by a high column density, high ionization 
absorber, where the sharp drop at 9 keV in the spectrum 
can be fitted by a high 
ionization (He or H-like) K-shell edge of iron. Both the column density 
and ionization of the absorber have to be rather extreme 
($N_{H}>3\times10^{23}$~cm$^{-2}$ and 
log~$\xi>3.5$~erg~cm$^{-1}$~s$^{-1}$ respectively), to produce a $\tau\sim1$ 
iron K-shell edge at 9~keV.  Indeed
\citet{Gallo04} have recently interpreted  
the strong Fe K edge in the \xmm\ X-ray spectra of 1H\,0707-495 
as due to a variable partial covering absorber and suggest that extreme disk 
reflection may not be required. Additionally, other high column 
density iron K-shell absorbers, with similar absorber parameters to 
\PG, have recently been observed in \xmm\ and {\it Chandra} observations of 
several high luminosity AGN; in particular PDS~456 \citep{R03}, 
PG~1211+143 \citep{Pounds03}, PG~1115+080 \citep{Chartas03} and 
APM~08279+5255 \citep{Chartas02,Hasinger02}. The outflow 
velocities inferred in these absorbers were generally found to be 
extreme, of up to 0.1c. Interestingly a {\it narrow} blue-shifted 
iron K$\alpha$ emission line at 7.5\,keV has been noted in the 
\asca\ spectrum of the NLS1, RX\,J0136.9-3510 \citep{Ghosh04}, 
these authors suggest that 
the line may also originate in an outflow with a velocity of $0.1-0.2c$. 
Indeed one possibility is that the absorber 
arises from the innermost part of an 
high velocity accretion disk outflow, within a few 
($10-100 R_{g}$) radii of the black hole. As noted by 
\citet{King03}, high velocity, large column density disk 
winds may be a natural consequence of accretion at a high fraction of the 
Eddington rate; in \PG\ the accretion rate is $\sim50\%$ of Eddington 
\citep{PROB04}. 

Determining which model (reflection vs absorber) is the most 
plausible explanation for the iron K profile in \PG\ is not possible 
with the current short \xmm\ observation. However a high column density 
absorber may impart 
discrete spectral features that could be detected with a longer 
observation (with \xmm) or with much improved spectral 
resolution (e.g. with the calorimeter-based XRS detector on 
Astro-E2, due for launch in 2005).  
For instance, resonant K-shell absorption lines 
from Fe~\textsc{xvii} through to Fe~\textsc{xxvi} may be produced 
in high ionization gas, while a Fe K$\beta$ unresolved transition 
array (UTA) at 7.2 keV may arise from moderately ionized matter 
\citep{Pal02}. Strong absorption lines may 
also be detected from other elements (e.g. Mg, Si and S),
making it possible to constrain the kinematics of the outflow. 
Indeed strong, 
blue-shifted (by up to 0.1c) absorption lines were detected in the 
XMM-Newton observations of PG~1211+143 \citep{Pounds03} and resonant 
iron absorption lines may even affect the iron line profiles of some nearby  
Seyfert 1 galaxies; e.g., NGC 3783 \citep{R04}. 
Alternatively if the spectrum of \PG\ can be 
explained by disk reflection, then it would be highly desirable to determine 
how the strong iron line changes with flux, in a much longer observation. 




\clearpage

\begin{figure}
\rotatebox{-90}{
\epsscale{0.7}
\plotone{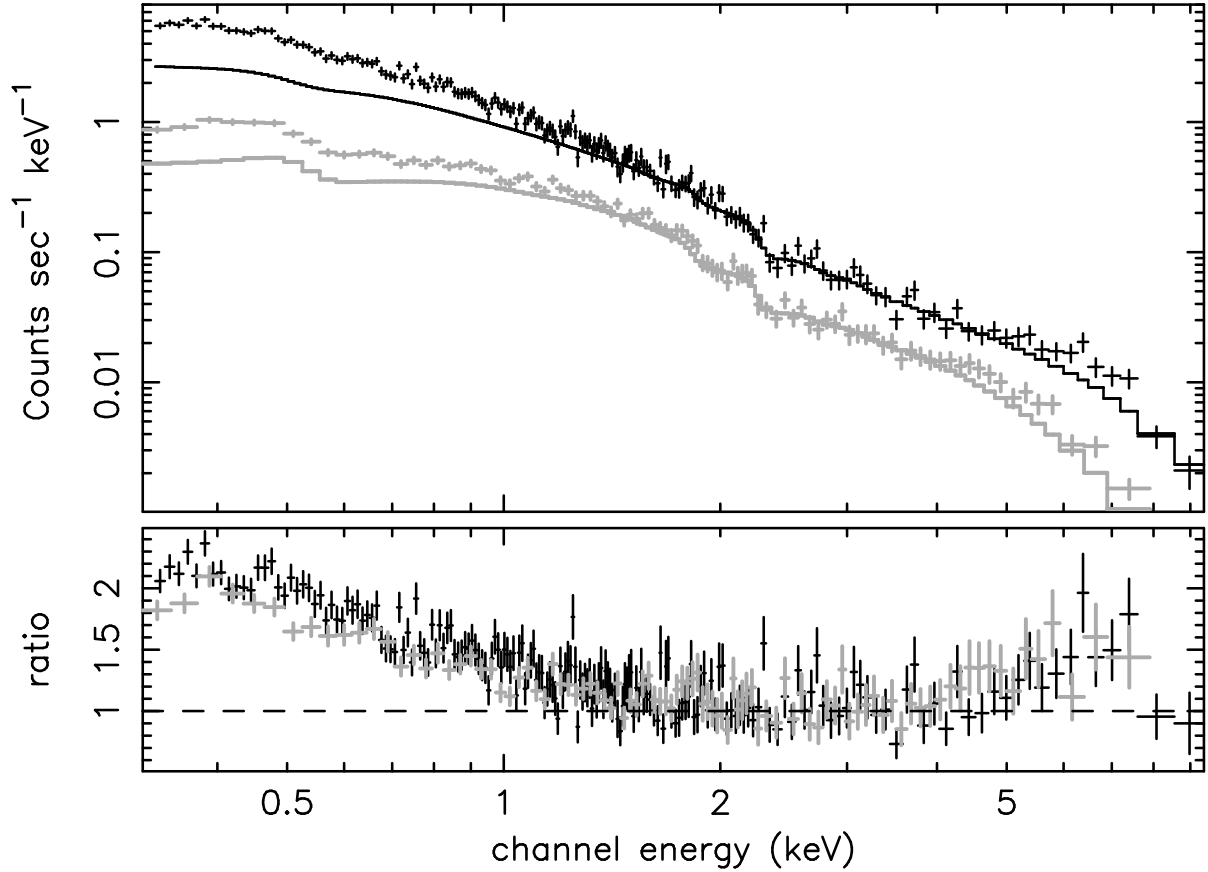}}
\caption{XMM-Newton EPIC-pn (black) and EPIC-MOS (grey) spectrum of \PG. 
Energy is plotted in the observed frame. 
The top panel shows the datapoints (crosses) and the model fitted to the 
data (solid line), folded through the detector response. The bottom panel 
shows the ratio of the data to a power-law continuum model, fitted between 
2-5 keV, of index $\Gamma=2.2$. A soft X-ray excess is clearly seen below 
1.5 keV, whilst an excess of counts is observed in both detectors above 5 keV 
in the iron K-shell band.}
\end{figure}

\begin{figure}
\rotatebox{-90}{
\epsscale{0.7}
\plotone{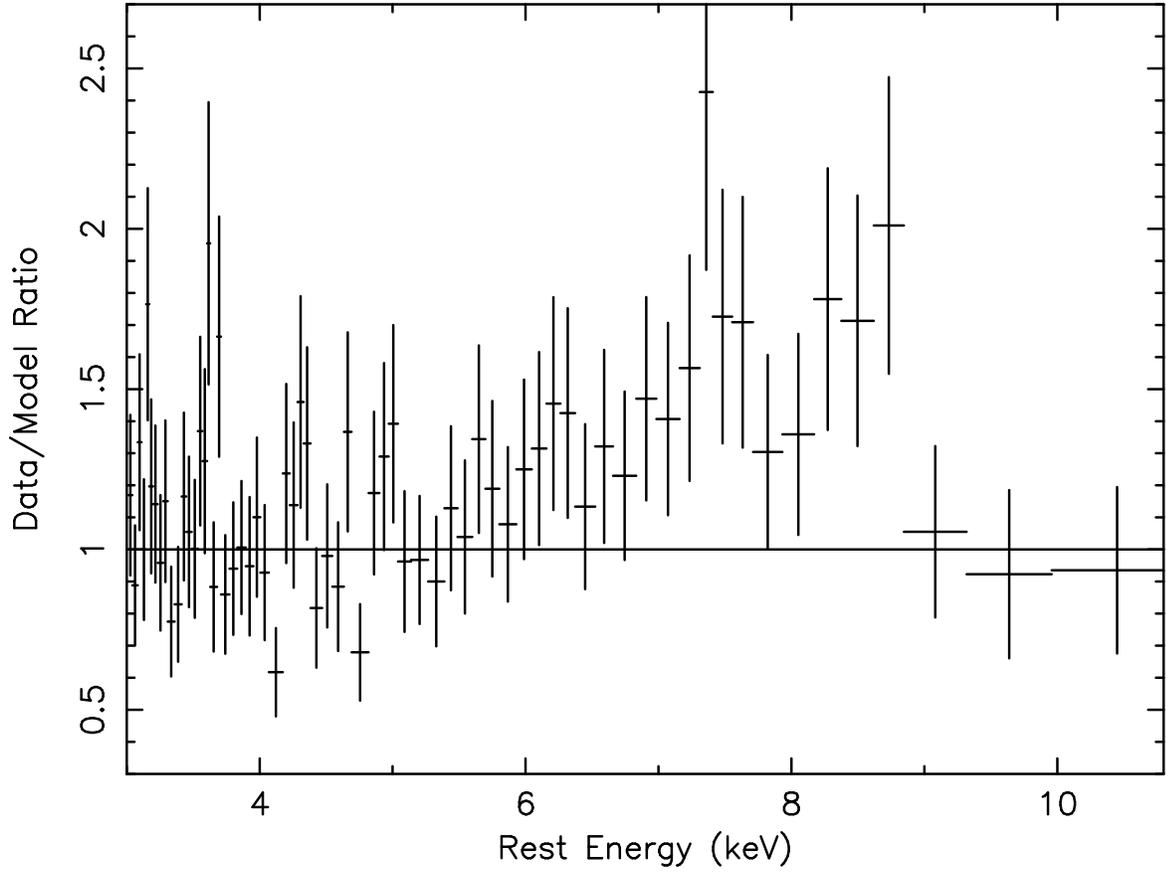}}
\caption{The iron line profile of \PG, showing the ratio of the EPIC-pn 
data to a 
broken power-law continuum fit, as described in the text. Energy is 
plotted in the quasar rest frame. An excess of counts, which can be modeled 
as a broad, but blue-shifted, iron K-shell line is observed between 6-9 keV, 
whilst a sharp drop is observed in the data at 9 keV.}
\end{figure}

\begin{figure}
\rotatebox{-90}{
\epsscale{0.7}
\plotone{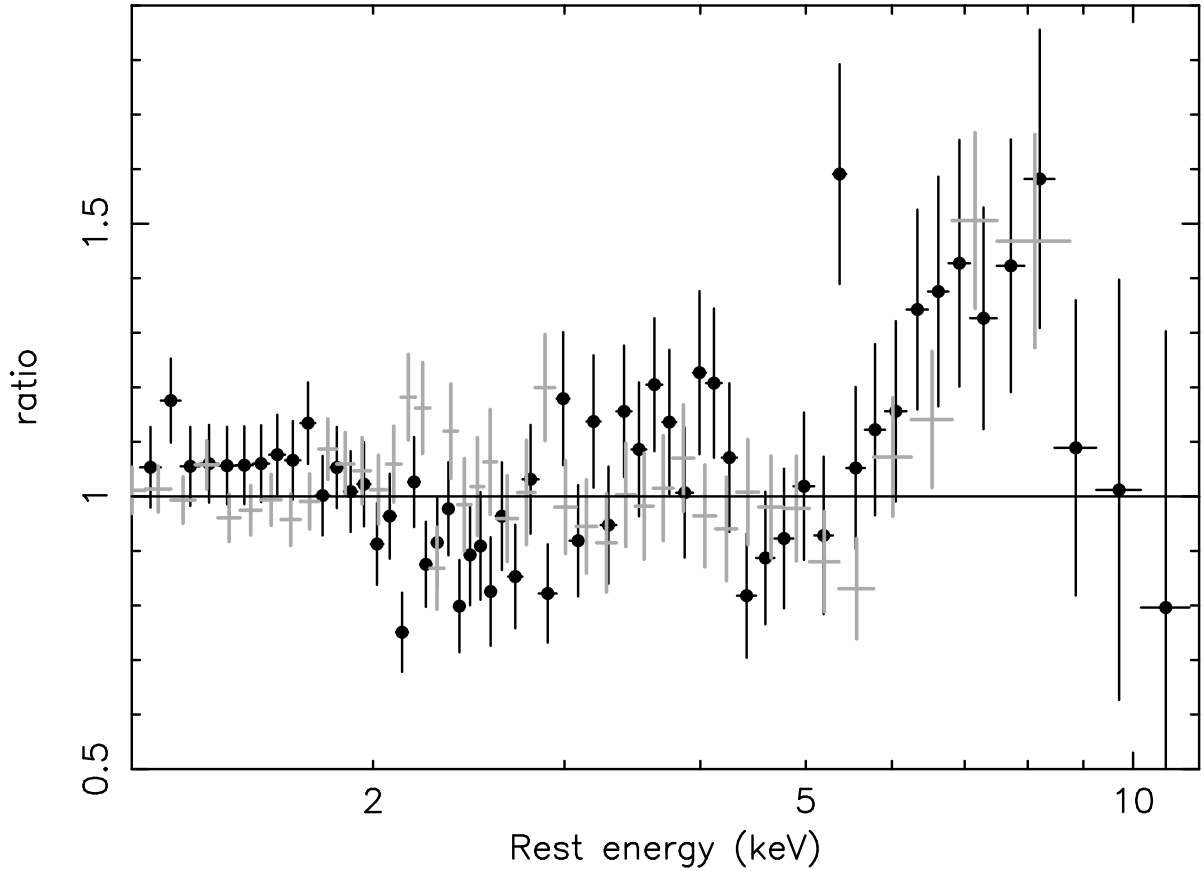}}
\caption{Data/Model ratio residuals to a simple power-law fit with 
Galactic absorption to the ASCA spectrum of \PG. The GIS data are shown 
as filled circles (black), the SIS data are crosses (grey). A strong 
and broad excess above 6 keV in the quasar rest frame, 
corresponding to the iron K-shell band, is 
clearly evident. Note however that the GIS2 and GIS3 data (and the 
SIS0 and SIS1 data) have been combined to improve the clarity of the 
display (but all 4 instruments have been fitted seperately 
in the ASCA spectral analysis.)}
\end{figure}

\begin{figure}
\rotatebox{-90}{
\epsscale{0.7}
\plotone{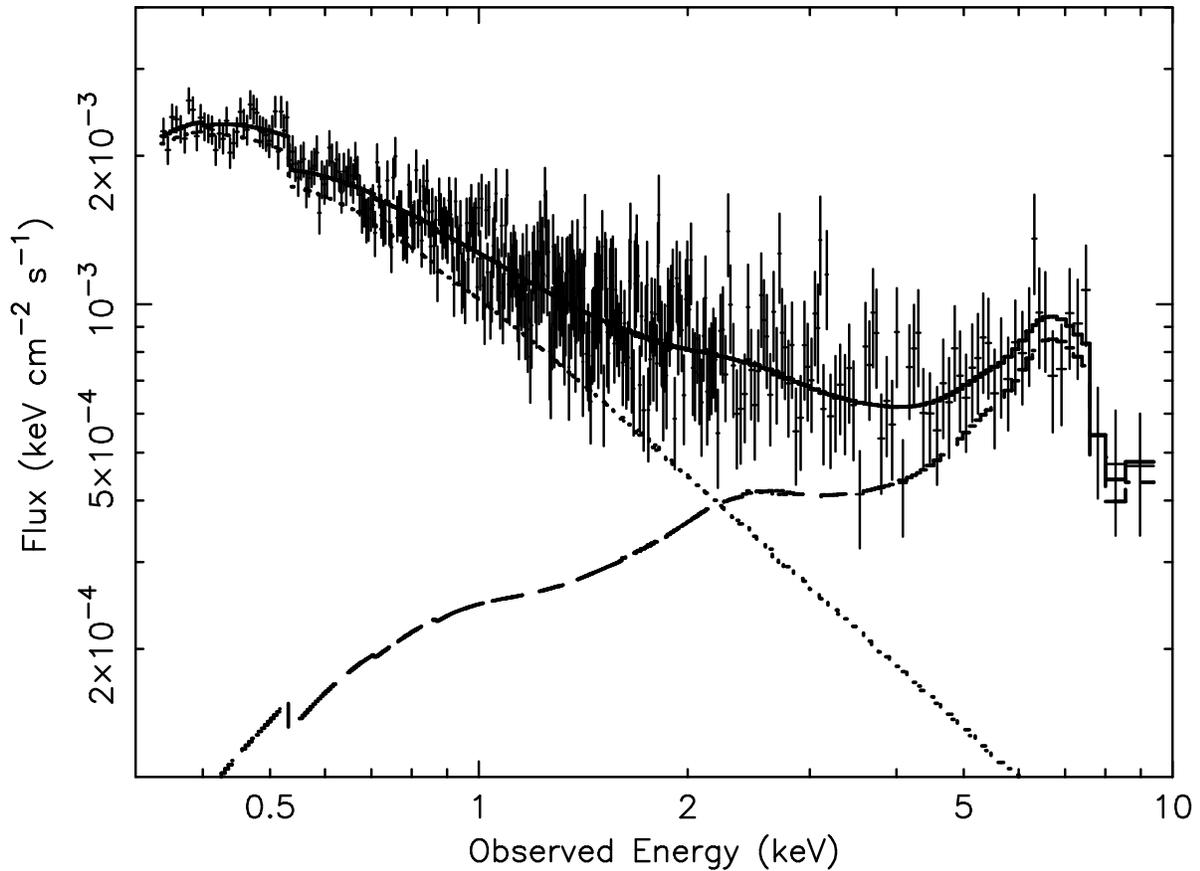}}
\caption{An ionized disc reflection model (\textsc{XION}), fitted to the 
spectrum of \PG. The crosses show the (unfolded) data-points, the solid line 
the total model emission, the dotted line the soft X-ray power-law and the 
dashed line the ionized reflection component. In order to 
model the extreme blue-wing of the line, the disc must be 
highly inclined at $>60\degg$. The reflection component 
must also dominate the emission in the iron K-band 
(i.e. the illuminating hard X-ray 
power-law is not directly observed), 
in order for the very strong iron line to be 
observed.}
\end{figure}

\begin{figure}
\rotatebox{-90}{
\epsscale{0.7}
\plotone{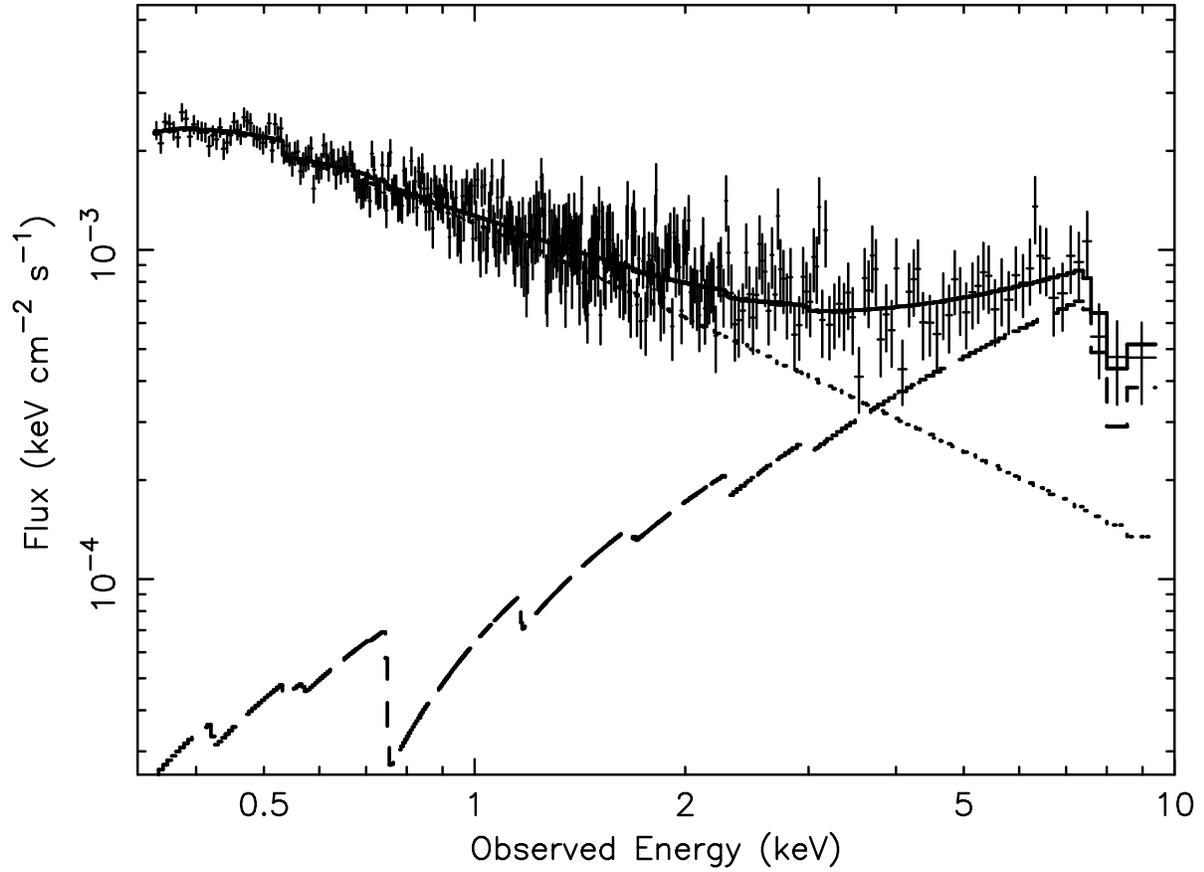}}
\caption{An ionized partial covering absorber model fit to the spectrum of 
\PG. The dotted line shows the unabsorbed soft X-ray power-law, whilst the 
dashed line shows the hard X-ray power-law, modified by the ionized absorber. 
A high column density of $>3\times10^{23}$~cm$^{-2}$ is required to model the 
deep absorption edge present in the spectrum of \PG\ at 9~keV.}
\end{figure}

\clearpage
\begin{deluxetable}{lccccccccc}\label{tab:line}
\tabletypesize{\small}
\tablecaption{Table of iron line spectral fits with relativistic line profiles.}
\tablewidth{0pt}
\tablehead{
\colhead{Fit} & \colhead{$\Gamma$\tablenotemark{a}} &  
\colhead{$E$\tablenotemark{b}} & \colhead{$\sigma$\tablenotemark{c} or 
$q$\tablenotemark{d} or $\theta$\tablenotemark{e}} & 
\colhead{$EW$\tablenotemark{b}} & \colhead{$R$\tablenotemark{g}} & 
\colhead{log\,$\xi$\tablenotemark{h}} & \colhead{$\chi^{2}/{\rm dof}$}} 

\startdata
{\sc Gaussian} & $2.47\pm0.08$ & 7.3$^{+0.5}_{-0.5}$ & 1.1$^{+0.9}_{-0.4}$ &  
1.5$^{+1.7}_{-0.7}$ & $1.6\pm0.7$ & $3.2\pm0.2$ & 397.2/377 \\

&\\
{\sc Gaussian}$^{i}$ & $2.4\pm0.2$ & 7.2$^{+0.6}_{-0.6}$ & $<1.4$ 
& 0.6$^{+0.5}_{-0.3}$ & 2(f) & 3.0(f) & 399.1/361 \\

&\\
{\sc laor} & $2.48\pm0.08$ & 8.5$^{+0.4}_{-0.3}$ & 3.5$^{+0.6}_{-0.4}$ 
& 3.9$^{+1.7}_{-1.3}$ & $1.5\pm0.4$ & $3.4\pm0.4$ & 394.2/377 \\

& & & 30\degg\ (f) \\

&\\
{\sc laor} & $2.46\pm0.08$ & 6.97(f) & 2.3$^{+0.6}_{-0.3}$ & 1.7$^{+0.7}_{-0.7}$ 
& $1.7\pm0.6$ & $3.2\pm0.3$ & 394.5/377 \\
&  &  & $68\pm5$\degg\ \\

\enddata


\tablenotetext{a}{Photon index of power-law continuum.}
\tablenotetext{b}{Rest frame energy or equivalent width of line in units of keV.}
\tablenotetext{c}{Velocity width ($\sigma$) in units of keV.}
\tablenotetext{d}{Emissivity power-law of the accretion disk.}
\tablenotetext{e}{Inclination of the accretion disk in degrees.}
\tablenotetext{f}{Parameter value is fixed.} 
\tablenotetext{g}{Reflection fraction (see text)}
\tablenotetext{h}{Log ionization parameter (see text)}
\tablenotetext{i}{Fit performed with ASCA data}

\end{deluxetable}


\end{document}